\documentclass[prl,aps,10pt,superscriptaddress,footinbib,twocolumn,floatfix,longbibliography]{revtex4-2}
\usepackage{amsmath,amssymb,amsfonts,amsthm,mathtools,bm}
\DeclareMathSizes{10}{10}{7}{5}
\usepackage{graphicx,microtype,tabularx,dcolumn,times}
\usepackage[unicode=true,bookmarks=true,bookmarksnumbered=false,
bookmarksopen=false,breaklinks=false,pdfborder={0 0 0},
backref=false,colorlinks=true,hyperfootnotes=false]{hyperref}
\hypersetup{linkcolor=magenta,urlcolor=blue,citecolor=blue,
pdfstartview={FitH}}
\usepackage{cleveref}
\newcommand{\ii}{\mathrm{i}}
\newcommand{\id}{\mathbf{1}}
\newcommand{\Tr}{\operatorname{Tr}}
\newcommand{\slD}{\gamma^b\nabla_b}
\newcommand{\cB}{\mathcal B}
\newcommand{\cM}{\mathcal M}
\newcommand{\cR}{\mathcal R}
\newcommand{\SM}{Supplemental Material}
\begin{document}
\title{Separation of Variables and Spectroscopy of Massive Spin-3/2 Fields on a Kerr Background}
\author{Ziqian Tang}
\email{tangziqian@pku.org.cn}
\affiliation{Beijing Key Laboratory of Fault-Tolerant Quantum Computing,
Beijing Academy of Quantum Information Sciences, Beijing 100193, China}
\affiliation{Beijing National Laboratory for Condensed Matter Physics,
Institute of Physics, Chinese Academy of Sciences, Beijing 100190, China}
\affiliation{University of Chinese Academy of Sciences, Beijing 100049, China}
\date{November 12, 2026}
\begin{abstract}
We separate the neutral massive Rarita--Schwinger equation on a Kerr
background at finite rotation. Two commuting matrix angular operators and a physical
first integral yield a second-order radial equation and reversible field
reconstruction. The resulting equations give quasibound spectra, greybody
factors, and helicity-resolved Hawking emission. For a matched weakly bound
level, the leading rotational splitting has the opposite sign to that of
a Dirac field. We derive its coefficient and identify its angular
momentum origin.
\end{abstract}
\maketitle

\paragraph{Introduction.}
Separation of variables reduces black-hole scattering and spectroscopy
to angular eigenvalue and radial boundary problems. On Kerr, the
Teukolsky equations separate massless scalar, neutrino, electromagnetic,
and gravitational fields \cite{Teukolsky1972,Teukolsky1973}, while
massless Rarita--Schwinger (RS) fields also admit separated descriptions
\cite{Guven1980,Kamran1985,TorresSilva1990}. Alternative separations of
the Maxwell potential use the hidden symmetry of rotating black-hole
geometries \cite{Lunin2017,Krtous2018}. For massive higher-spin fields,
additional polarizations and constraints obstruct a direct extension of
the massless description. Separated equations are nevertheless known
for massive scalar fields \cite{CarterSeparation1968,Brill1972} and Dirac
fields \cite{Chandrasekhar1976,PageDirac1976,CarterMcLenaghan1979}, and
for all three Proca polarizations \cite{Frolov2018,DolanProca2018}.
More recently, Dias and Santos separated massive spin-2 fields on Kerr
using commuting matrix angular operators, radial transport, and a
physical first integral \cite{DiasSantos2026}. For the massive spin-$3/2$
RS field, Schwarzschild separation \cite{Zecca2006,Zecca2007} provides
a spherical starting point. At finite Kerr rotation, separation must
account for angular coupling and the vector--spinor constraints while
retaining all physical states and reconstructing the full field.

The spin-$3/2$ vector--spinor described by the RS equation
\cite{Rarita1941} appears in the gravitino sector of suitable supergravity
theories. A cosmologically stable, minimally coupled massive spin-$3/2$
particle is also a dark-matter candidate that can be populated by
gravitational production during and after inflation
\cite{KolbRaritron2026}. Such cosmological production can depend strongly
on the longitudinal helicity-$1/2$ sector
\cite{Kolb2021,KolbRaritron2026}. Black-hole emission provides a
complementary setting in which both massive propagation and the
physical helicity content determine the production spectrum.
Consistent propagation depends on the background and the couplings
\cite{Frauendiener2003,Schenkel2012}.

In this work, we consider a neutral massive RS test field with minimal
gravitational coupling and constant nonzero mass on the Ricci-flat Kerr
background. We construct its commuting angular operators, spectral
relation, and reversible reconstruction. A joint angular
eigenfunction fixes the relative component profiles and leaves one
physical second-order radial problem. The separated equations yield
quasibound frequencies, greybody factors, and helicity-resolved Hawking
emission. At matched quantum numbers, the leading rotational splitting
has the opposite sign to that of a massive Dirac field
\cite{DolanDempsey2015}.

\paragraph{Field equations.}
Let $\psi_a$ be a covector--spinor of mass $\mu$, with orthonormal-frame
indices $a,b,c=0,1,2,3$. We use $\eta^{ab}=\operatorname{diag}(-1,1,1,1)$
and Dirac matrices $\{\gamma^a,\gamma^b\}=2\eta^{ab}$;
$\gamma^{ab}$ and $\gamma^{abc}$ denote unit-weight antisymmetrized
products. Repeated frame indices are contracted with $\eta^{ab}$.
Braces and brackets denote anticommutators and commutators,
respectively, and $\ii^2=-1$. The covariant derivative $\nabla_b$ acts
on both field indices. The RS operator $\mathcal E^a$ is defined by
\begin{equation}
 \mathcal E^a[\psi]\equiv
 \gamma^{abc}\nabla_b\psi_c-\mu\gamma^{ac}\psi_c=0 .
 \label{eq:RS}
\end{equation}
For a constant $\mu\ne0$
on a Ricci-flat background, its gamma contraction and divergence imply
\begin{equation}
 (\slD+\mu)\psi_a=0,\qquad
 \gamma^a\psi_a=0,\qquad \nabla^a\psi_a=0 .
 \label{eq:constraints}
\end{equation}
These equations retain the four massive particle spin states. We take
$\mu>0$ and use $G=c=\hbar=1$, with $G$ Newton's constant, $c$ the
speed of light, and $\hbar$ the reduced Planck constant.

Here $M>0$ is the black-hole mass and $a$ its angular momentum per unit
mass, with $|a|\le M$. Let $\Delta=r^2-2Mr+a^2$ and
$\Sigma=r^2+a^2\cos^2\theta$. The ingoing
coordinates $(T,r,\theta,\varphi)$ obey
$dT=dt+2Mr\,dr/\Delta$ and
$d\varphi=d\phi+a\,dr/\Delta$ relative to Boyer--Lindquist coordinates
$(t,r,\theta,\phi)$, where $T,t$ are time coordinates, $r$ is radius,
$\theta$ the polar angle, and $\varphi,\phi$ the respective azimuths.
Stationary modes have dependence $e^{-\ii\omega T+\ii m\varphi}$:
$\omega_R,\omega_I$ are the real and imaginary parts of the frequency
$\omega=\omega_R+\ii\omega_I$, and $m$ the half-integer
azimuthal angular momentum. The dimensionless parameters are
$\alpha=M\mu$, $\chi=a/M$, and $\widehat\omega=M\omega$;
$\widehat\omega_R=M\omega_R$ and $\widehat\omega_I=M\omega_I$.

\paragraph{Separation and reconstruction.}
Choose a principal orthonormal frame parallel along
$\partial_r-\partial_T$. Two highest-null-weight projections of $\psi_a$
and two of its gamma-traceless curl define four master variables.
After a constant mass-dependent normalization, write them as
$\bm\Omega=(\Omega_{3/2},\Omega_{1/2},\Omega_{-1/2},\Omega_{-3/2})^{\mathsf T}$.
Here $\mathsf T$ denotes transpose, and the subscripts are spin weights,
the rotation-generator eigenvalues of the local angular frame.
Extraction is specified in
Sec.~S1.2 of the \SM.

For $y=\cos\theta$, $W=(1-y^2)^{1/2}$, and the standard spin-$3/2$ matrices
$J_i$ ($i=x,y,z$), with
$J_z=\operatorname{diag}(3/2,1/2,-1/2,-3/2)$, define the angular
operators $K_1,K_2$ below. The identity $\id$ has the dimension of the
component space on which it acts:
\begin{equation}
\begin{aligned}
 K_1={}&\frac{W}{\mu}J_y\partial_y-\frac{yJ_y}{2\mu W}
 +\frac{\ii}{\mu}\left(\frac mW-a\omega W\right)J_x
 \\
 &-\frac{\ii y}{2\mu W}\{J_x,J_z\}
 -\ii ayJ_z+\frac{\ii\omega}{\mu^2}J_z^2 ,
\end{aligned}
 \label{eq:K1}
\end{equation}
\begin{equation}
\begin{aligned}
 K_2={}&\partial_y[(1-y^2)\partial_y]
 -\frac{m^2\id-2myJ_z+J_z^2}{1-y^2}
 \\
 &+a^2(\omega^2-\mu^2)y^2\id
 +2a\omega yJ_z+2a\mu WJ_x .
\end{aligned}
 \label{eq:K2}
\end{equation}
The operators are independent of $r$ and $M$ and satisfy exact differential identities
$[K_1,K_2]=0$ and $F(K_1,K_2)=0$. The polynomial $F$ is quartic in its
first spectral argument; its coefficients are given in Sec.~S1.2.
Writing $\bm S$ for a regular four-component angular profile,
$\lambda,\nu$ for its joint eigenvalues, and $f$ for the radial amplitude,
we have
\begin{equation}
 K_1\bm S=\lambda\bm S,\quad K_2\bm S=\nu\bm S,\quad
 \bm\Omega=e^{-\ii\omega T+\ii m\varphi}f(r)\bm S(\theta).
 \label{eq:ansatz}
\end{equation}
The relation $F(\lambda,\nu)=0$ ties the two eigenvalues together.

Define the radial differential operator $\cB$ and its scalar
coefficients $B,V$ by
\begin{equation}
\begin{aligned}
 \cB&=\Delta\partial_r^2+B\partial_r+V,\\
 B&=5(r-M)-4\ii Mr\omega+2\ii am,\\
 V&=(\omega^2-\mu^2)r^2+2M\omega^2r
       -3\ii\omega r-5\ii M\omega+\frac{15}{4}.
\end{aligned}
 \label{eq:B}
\end{equation}
The original equations imply the closed radial transport relation
\begin{equation}
 L_3f\equiv(\partial_r+\ii\omega)[(\cB+\nu)f]
 +(2\mu^2\lambda-3\mu^2r-\frac92\ii\omega)f=0 .
 \label{eq:L3}
\end{equation}
A cubic polynomial $A(r)$, with leading coefficient $\mu^3$, supplies the
physical first integral; primes below denote $r$ derivatives:
\begin{equation}
\begin{aligned}
 L_2f={}&A(\cB+\nu)f-\Delta A'f'\\
 &+\{\Delta A''+(\Delta'-B+\ii\omega\Delta)A'\}f=0 .
\end{aligned}
 \label{eq:L2}
\end{equation}
On the spectral curve,
$(\partial_r+\ii\omega)L_2=A L_3$.
Thus transport alone permits $L_2f=C_{\rm int} e^{-\ii\omega r}$,
where $C_{\rm int}$ is an integration constant.
The original RS equations impose $C_{\rm int}=0$. Equation~\eqref{eq:L2} is
second order wherever $A\ne0$; at its zeros the joint system
$L_2f=L_3f=0$ retains the physical continuation. The explicit $A$ and the
derivation are given in Sec.~S1.3.

The reconstruction is an essential part of this reduction. Let $\cM$
extract the master variables and $\cR$ reconstruct the physical field.
These differential maps satisfy
\begin{equation}
 \cM\cR\bm\Omega=\bm\Omega,\qquad
 \cR\cM\psi=\psi
 \label{eq:inverse}
\end{equation}
on the respective solution spaces. To see why no field is lost, set its
four masters to zero. Radial transport reduces the remaining components
to four auxiliary angular functions $h_j$, $j=2,3,4,5$.
Two polynomial identities from the
original equations force $\mu^2h_2=\mu^2h_5=0$, followed by
$\mu h_3=\mu h_4=0$. The remaining field components then vanish.
The same elimination with nonzero masters constructs the inverse;
Sec.~S1.4 gives the finite construction directly from the field equations.

An off-axis representative has order four in $(r,\theta)$ at fixed
$(\omega,m)$. An equivalent representative of total differential order six
extends smoothly to both axes and the future horizon. Its coefficients
contain no inverse horizon function or spectral parameter.
For real $\omega$, the self-adjoint elliptic $K_2$ and the commuting
anti-self-adjoint $K_1$ give a complete joint angular expansion.
For complex $\omega$ we use individual joint modes, without asserting
completeness of the complex-frequency spectrum.

\begin{figure}[tbp]
 \centering
 \makebox[\columnwidth][c]{\hspace*{-0.12in}\includegraphics[width=\columnwidth]{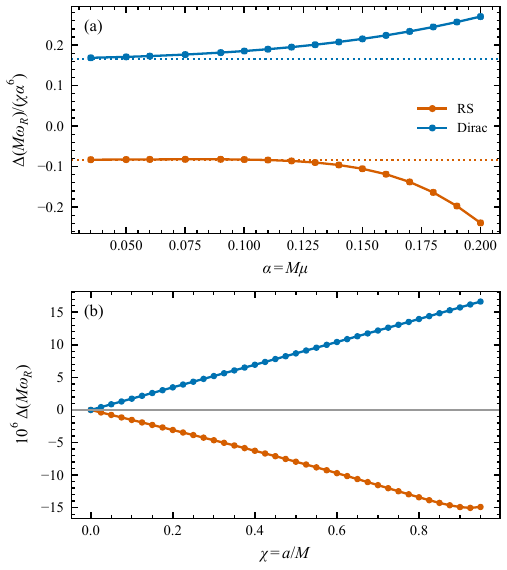}\hspace*{0.12in}}
 \caption{Rotational ordering at matched principal number $n=2$,
 orbital angular momentum $\ell=1$, and total angular momentum $j=1/2$.
 (a) $\Delta\widehat\omega_R/(\chi\alpha^6)$, where
 $\Delta\widehat\omega_R$ is the $m=+1/2$ minus $m=-1/2$ difference.
 Results use
 $\chi=0.01$ (circles) and $0.02$ (crosses); dotted lines are the
 leading coefficients in Eq.~\eqref{eq:splitting}.
 (b) Finite-rotation splitting at $\alpha=0.2$, with $\chi$ sampled
 in steps of $0.025$.
 Lines join computed frequencies.}
 \label{fig:splitting}
\end{figure}

\paragraph{Quasibound spectroscopy.}
We impose the ingoing regular solution at the future horizon and decay at
infinity, choosing the inverse decay length
$q=\sqrt{\mu^2-\omega^2}$ with $\operatorname{Re}q>0$.
The frequency is complex. With our time convention, $\omega_I<0$ gives
decay of the amplitude on a time $-1/\omega_I$. The occupation decay
time $\tau_N$ is
\begin{equation}
 \tau_N/M=[-2\operatorname{Im}\widehat\omega]^{-1}.
 \label{eq:tau}
\end{equation}
A non-Hermitian joint angular solve is included at every trial frequency.
The principal, orbital, and total angular momentum labels $(n,\ell,j)$
in Schwarzschild identify branches continued to
finite rotation; only $m$ remains a conserved angular quantum number there.

The leading weak-binding radial equation is hydrogenic. Its principal
quantum number is $n=n_r+\ell+1$, where $n_r$ is the nonnegative radial
overtone number, with
$\omega_R/\mu=1-\alpha^2/(2n^2)+\cdots$.
For example, the two $j=1/2$ RS branches have $\ell=1,2$; the corresponding
Dirac branches have $\ell=0,1$. To compare rotation without changing
orbital or principal labels, we select $n=2$, $\ell=1$, $j=1/2$ in both
theories and define
$\Delta\widehat\omega_R=\widehat\omega_R(+1/2)-\widehat\omega_R(-1/2)$.
The leading slow-rotation coefficients are
\begin{equation}
 \lim_{\alpha\to0}\alpha^{-6}
 \left.\partial_\chi\Delta\widehat\omega_R\right|_{\chi=0}
 =
 \begin{cases}
 -1/12,&\text{massive RS},\\
 +1/6,&\text{Dirac}.
 \end{cases}
 \label{eq:splitting}
\end{equation}
The weakly bound RS level has the opposite rotational ordering.

This coefficient follows directly from the separated equations.
For $M=1$, introduce the scaled radius $\rho=\mu^2r$ and
$\omega=\mu+\mu^3e+a\mu^6c$, where $e$ is the leading binding-energy
coefficient and $c$ the rotational frequency coefficient.
After removing the first radial derivative,
the unperturbed reduced wavefunction is
$U_0=\rho^2e^{-\rho/2}$, with $e=-1/8$. The coefficient $\delta V$
of $a\mu^3$ in its normal-form potential is
\begin{equation}
 \delta V=2c-\frac{16em}{3\rho^2}
              -\frac{4m}{\rho^3}+\frac{20m}{3\rho^4}.
 \label{eq:pert}
\end{equation}
The condition $\int_0^\infty U_0^2\delta V\,d\rho=0$ gives $c=-m/12$.
Repeating the calculation for the Dirac equations gives $c=m/6$
(Sec.~S2.3). These are leading exterior weak-binding coefficients.

The sign also has an angular interpretation. In the standard leading
frame-dragging Hamiltonian \cite{Obukhov2009}, let $\bm L$ be orbital
angular momentum, $\bm{\mathsf S}$ intrinsic spin, and $\bm n$ the radial
unit vector, with the $z$ axis along the black-hole spin:
\begin{equation}
 H_J=\frac{aM}{r^3}
 [2L_z+3n_z(\bm n\cdot\bm{\mathsf S})-\mathsf S_z],
 \label{eq:HJ}
\end{equation}
The brackets $\langle\cdot\rangle$ denote expectation values in the
specified bound state. The $\ell=1$, $j=m=1/2$ state has
$\langle L_z\rangle=2/3$ for intrinsic spin $1/2$ and $-1/3$ for intrinsic
spin $3/2$. The complete angular bracket in Eq.~\eqref{eq:HJ} is $2$
and $-1$, respectively. Together with
$\langle r^{-3}\rangle=\mu^6/24$ at $M=1$, these matrix elements reproduce
Eq.~\eqref{eq:splitting}. The sign follows from the different angular
content of the two fields.

Fig.~\ref{fig:splitting} compares the asymptotic coefficients with
numerical frequencies and shows the ordering at finite rotation.
We also follow four selected families at $\alpha=0.2$, with
$m=\pm1/2$ or $\pm3/2$, through $\chi=0.95$. Their lifetimes span several
orders of magnitude. One $j=1/2$ co-rotating mode lives about twenty times
longer at $\chi=0.9$ than at $\chi=0$, whereas two $j=3/2$ families have
local lifetime maxima near $\chi=0.6$. Our matched Dirac calculation also
has lifetime extrema, so these extrema do not distinguish the RS field.
All followed modes decay; this finite set does not
establish mode stability.

Selected frequencies and occupation decay times are given in Sec.~S2.2.

\begin{figure}[tbp]
 \centering
 \makebox[\columnwidth][c]{\hspace*{-0.10in}\includegraphics[width=\columnwidth]{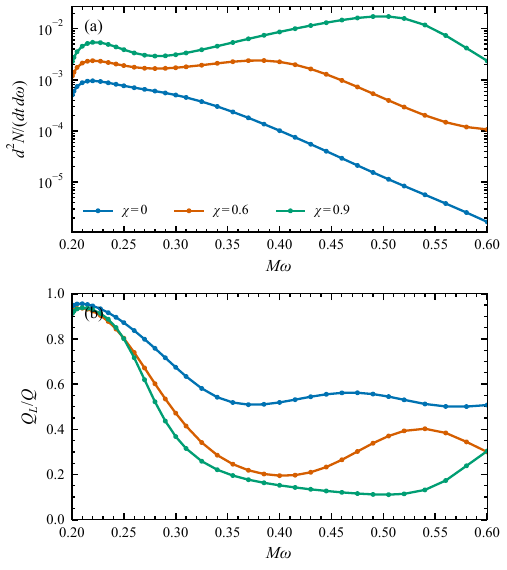}\hspace*{0.10in}}
 \caption{Emission at $\alpha=0.2$ for $\chi=0$, $0.6$, and $0.9$.
 (a) Particle spectra with the standard fermionic Unruh occupation.
 (b) $Q_L/Q$, where $Q$ is the total number spectrum and $Q_L$ its
 contribution from asymptotic states of helicity $h=\pm1/2$.
 The displayed sums include
 the branches continued from Schwarzschild $j\le7/2$.
 Dots are computed frequencies; lines guide the eye.}
 \label{fig:emission}
\end{figure}

\paragraph{Greybody factors and emission.}
We normalize scattering using the conserved bilinear current
\cite{Schenkel2012}
\begin{equation}
 j^a=-\ii\bar\psi_b\gamma^{bac}\psi_c,\qquad
 \Phi=2\pi\int_{-1}^{1}\Sigma j^r\,dy ,
 \label{eq:flux}
\end{equation}
Here $j^a$ is the frame current, $j^r$ its radial coordinate component,
$\Phi$ its outward radial flux, and
$\bar\psi=\psi^\dagger(-\ii\gamma^0)$, with $\dagger$ denoting Hermitian
conjugation.
For a real-frequency channel $\omega>\mu$, the greybody factor is
$\Gamma=-\Phi_H/\Phi_{\rm in}$, where $\Phi_H$ is the outward horizon
flux and $\Phi_{\rm in}>0$ the incident flux magnitude.
We compute the incoming and outgoing flux matrices before forming the
unit-flux reflection matrix $S_m$, which maps incoming to outgoing
channels at fixed $m$. This avoids identifying squared master
amplitudes with physical probabilities.

Let $r_+$ be the outer horizon radius, $\Omega_H$ its angular velocity,
and $T_H$ its Hawking temperature. For the standard free-fermion Unruh
occupation \cite{Page1976}, with $\Omega_H=a/(2Mr_+)$,
$T_H=\sqrt{M^2-a^2}/(4\pi Mr_+)$, and
$r_+=M+\sqrt{M^2-a^2}$, the particle spectrum is
\begin{equation}
 Q(\omega)=\frac{d^2N}{dt\,d\omega}
 =\frac1{2\pi}\sum_m
 \frac{\Tr(\id-S_mS_m^\dagger)}
 {e^{(\omega-m\Omega_H)/T_H}+1}.
 \label{eq:emission}
\end{equation}
Here $N$ is the emitted particle number and $\Tr$ the trace over
physical angular channels. The mode sum includes all four
massive particle spin states, for one Majorana species or one charge
sector of a Dirac field. Including particles and antiparticles
of a Dirac species doubles the rate.

At infinity, helicity $h$ is the eigenvalue of the total radial spin
generator $\mathsf H$ defined in Sec.~S3.2. The projector
$P_L=(9\id/4-\mathsf H^2)/2$ selects longitudinal states $|h|=1/2$;
$P_T=\id-P_L$ selects transverse states $|h|=3/2$.
In the outgoing unit-flux basis, $P_L$ is represented by $\mathsf L_m$.
Replacing the numerator of
Eq.~\eqref{eq:emission} by
$\Tr[\mathsf L_m(\id-S_mS_m^\dagger)]$ defines the longitudinal
particle spectrum $Q_L$.

The helicity content changes across the spectrum
(Fig.~\ref{fig:emission}). On the common band
$0.2005\le M\omega\le0.6$, the longitudinal particle fractions,
$\int Q_L\,d\omega/\int Q\,d\omega$ over that band, are
$0.776$, $0.444$, and $0.220$ for $\chi=0$, $0.6$, and $0.9$.
Reweighting fixed-$m$ transmission traces shows that both the occupation
factor and the helicity content of the transmitted channels matter.
Massless greybody factors with a kinematic mass cutoff
\cite{BlackHawk2021,BlackHawk2026} do not in general reproduce this finite-mass
spectrum. The comparison and retained frequency sums are given in
Sec.~S3.3. We assume the usual physical-mode Unruh prescription;
the calculation does not establish global quantization of the
constrained Kerr field.

\paragraph{Discussion.}
The massive RS equations on Kerr admit a matrix angular separation with a
physical second-order radial problem and reversible field reconstruction.
This structure permits both resonance calculations and flux-based
scattering. The matched hyperfine coefficient gives a simple spectral
distinction from a Dirac fermion, while the emission calculation resolves
the mass-allowed longitudinal states at infinity. The separation is at
finite rotation; the analytic splitting is a weak-binding,
slow-rotation result, and the spectra sample specified mode families.
These equations provide a basis for further searches of the massive
spin-$3/2$ resonance spectrum and for particle-production calculations
with controlled physical mode counting.

\nocite{Kerr1963,Carter1968,Michishita2019,Hawking1975,Calza2026}
\bibliography{ref}
\end{document}


\setcounter{secnumdepth}{3}
\renewcommand{\thesection}{S\arabic{section}}
\renewcommand{\thesubsection}{S\arabic{section}.\arabic{subsection}}
\renewcommand{\thefigure}{S\arabic{figure}}
\renewcommand{\thetable}{S\arabic{table}}
\makeatletter
\def\@seccntformat#1{\csname the#1\endcsname\quad}
\renewcommand{\p@subsection}{}
\def\frontmatter@title@format{\large\normalfont\bfseries\centering\parskip\z@skip}
\makeatother
\title{Supplemental Material for\texorpdfstring{\\}{ }
``Separation of Variables and Spectroscopy of Massive Spin-3/2 Fields on a Kerr Background''}
\author{Ziqian Tang}
\email{tangziqian@pku.org.cn}
\affiliation{Beijing Key Laboratory of Fault-Tolerant Quantum Computing,
Beijing Academy of Quantum Information Sciences, Beijing 100193, China}
\affiliation{Beijing National Laboratory for Condensed Matter Physics,
Institute of Physics, Chinese Academy of Sciences, Beijing 100190, China}
\affiliation{University of Chinese Academy of Sciences, Beijing 100049, China}
\date{November 12, 2026}
\maketitle

This Supplemental Material has three sections. Section~S1 gives the
field conventions, separated equations, and reversible reconstruction.
Section~S2 describes quasibound states and derives the rotational
splitting relative to the Dirac field. Section~S3 presents flux
normalization, helicity resolution, and Hawking emission.
Units and conventions are specified below.

\section{Separation and reversible reconstruction}
\label{sec:conventions}
\subsection{Background and constraints}
We set Newton's constant $G$, the speed of light $c$, and the reduced
Planck constant $\hbar$ to one. The background is the four-dimensional Kerr vacuum
\cite{Kerr1963,Carter1968}, where $M>0$ is the black-hole mass and $a$ its
angular momentum per unit mass, with $|a|\le M$, horizon radii
$r_\pm=M\pm\sqrt{M^2-a^2}$ (outer and inner, respectively), and
\begin{equation}
 \Delta=r^2-2Mr+a^2,\quad \Sigma=r^2+a^2y^2,\quad
 y=\cos\theta,\quad W=\sqrt{1-y^2}.
 \label{eq:geometry}
\end{equation}
Here $r$ is radius and $\theta$ the polar angle. The ingoing
Kerr--Schild (KS) coordinates $(T,r,\theta,\varphi)$ and
Boyer--Lindquist (BL) coordinates $(t,r,\theta,\phi)$ have time
coordinates $T,t$ and azimuths $\varphi,\phi$, related by
\begin{equation}
 \dd T=\dd t+\frac{2Mr}{\Delta}\dd r,\qquad
 \dd\varphi=\dd\phi+\frac{a}{\Delta}\dd r .
 \label{eq:coordinates}
\end{equation}
In $(T,r,\theta,\varphi)$ the metric line element $\dd s^2$ is
\begin{equation}
\begin{aligned}
 \dd s^2={}&-\dd T^2+\dd r^2+\Sigma\dd\theta^2
 +(r^2+a^2)W^2\dd\varphi^2-2aW^2\dd r\,\dd\varphi\\
 &+\frac{2Mr}{\Sigma}
   (\dd T+\dd r-aW^2\dd\varphi)^2 .
\end{aligned}
 \label{eq:metric}
\end{equation}
Here $g_{\alpha\beta}$ are the coordinate metric components and
$g=\det g_{\alpha\beta}$ their determinant, with
$\alpha,\beta$ denoting coordinate indices. Its volume density is
$\sqrt{-g}=\Sigma\sin\theta$.
Let $E_A=E_A{}^\alpha\partial_\alpha$ be the orthonormal frame, with
frame indices $A,B,C=0,1,2,3$. In coordinate order $(T,r,y,\varphi)$,
\begin{equation}
 E_A{}^\alpha=
 \begin{pmatrix}
 1+H_{\rm KS}&-H_{\rm KS}&0&0\\ H_{\rm KS}&1-H_{\rm KS}&0&0\\
 0&-a^2yW/\Sigma&-rW/\Sigma&-ay/(W\Sigma)\\
 0&arW/\Sigma&-ayW/\Sigma&r/(W\Sigma)
 \end{pmatrix},\qquad H_{\rm KS}=\frac{Mr}{\Sigma}.
 \label{eq:tetrad}
\end{equation}
Thus $E_1-E_0=\partial_r-\partial_T$.
The full vector and spin connections along this direction vanish.
At the axes, the polar frame is replaced by smooth spin-frame patches.

We use the frame metric $\eta_{AB}=\diag(-1,1,1,1)$ and
Dirac matrices $\gamma^A=\ii\gamma_{\rm std}^A$, where
$\gamma_{\rm std}^A$ is the usual Dirac representation in signature
$(+---)$ and $\ii^2=-1$. Thus $\{\gamma^A,\gamma^B\}=2\eta^{AB}$;
$\gamma^{AB}=\gamma^{[A}\gamma^{B]}$ and higher products have
unit-weight antisymmetrization; brackets on indices denote this operation.
Repeated frame indices are contracted with $\eta^{AB}$.
Braces and operator brackets denote
anticommutators and commutators; $\id_d$ is the $d$-dimensional identity,
and $\id$ has the dimension fixed by its operand.
For a covector--spinor $\psi_A$, the derivative is
\begin{equation}
 \nabla_\alpha\psi_A=\partial_\alpha\psi_A+
 \frac14\omega_{\alpha BC}\gamma^{BC}\psi_A
 -\omega_\alpha{}^B{}_A\psi_B .
 \label{eq:covder}
\end{equation}
Here $\omega_{\alpha BC}$ is the frame connection fixed by the tetrad
and the torsion-free Cartan equation; it is distinct from the mode
frequency $\omega$. Lower-case Latin indices $a,b,c$ also denote frame
indices. We write $\slD$ for the covariant Dirac operator.

Set $\beta=-\ii\gamma^0$ and $\bar\psi_A=\psi_A^\dagger\beta$, where
$\dagger$ is Hermitian conjugation. The action of the neutral
Rarita--Schwinger (RS) field of constant mass $\mu>0$ is
\begin{equation}
 \mathcal I=\int\dd^4x\sqrt{-g}\left[
 -\frac12\bar\psi_a\gamma^{abc}
       \overleftrightarrow{\nabla}_b\psi_c
 +\mu\bar\psi_a\gamma^{ac}\psi_c\right].
 \label{eq:action}
\end{equation}
Here $x$ denotes the spacetime coordinates and
$\bar\psi\overleftrightarrow\nabla_b\psi
=\bar\psi\nabla_b\psi-(\nabla_b\bar\psi)\psi$.
Variation gives the field operator
$\mathcal E^a[\psi]=\gamma^{abc}\nabla_b\psi_c-\mu\gamma^{ac}\psi_c$.
Let $\vartheta=\gamma^a\psi_a$ be its gamma trace. Gamma contraction of
$\mathcal E^a[\psi]=0$
in four dimensions gives
\begin{equation}
 2(\slD\vartheta-\nabla^a\psi_a)-3\mu\vartheta=0.
 \label{eq:gamconstraint}
\end{equation}
The divergence of the RS operator contains a curvature commutator and
$-\mu(\slD\vartheta-\nabla^a\psi_a)$. The contracted curvature term
vanishes on a Ricci-flat background. Hence $\mu^2\vartheta=0$;
for $\mu\ne0$, $\vartheta=\nabla^a\psi_a=0$.
Expanding $\gamma^{abc}$ reduces the equation to
$(\slD+\mu)\psi_a=0$. Conversely these constrained Dirac-type equations
give the original RS equation. The background and nonzero mass are
part of this equivalence \cite{Frauendiener2003,Schenkel2012}.

Stack the four spinor blocks in order $(\psi_0,\psi_1,\psi_2,\psi_3)$.
The independent components are
$u=(\psi_1,\psi_2,\psi_3)^{\mathsf T}\in\mathbb C^{12}$,
where $\mathsf T$ denotes transpose; the embedding $\psi=Nu$ is
\begin{equation}
 N=\begin{pmatrix}
 \gamma^0\gamma^1&\gamma^0\gamma^2&\gamma^0\gamma^3\\
 \id&0&0\\0&\id&0\\0&0&\id
 \end{pmatrix}.
 \label{eq:embedding}
\end{equation}
The divergence is retained in the differential system. Majorana reality
changes species counting, while the local equations are the same.
Modes have phase $e^{-\ii\omega T+\ii m\varphi}$, where
$\omega_R,\omega_I$ are the real and imaginary parts of the mode
frequency $\omega=\omega_R+\ii\omega_I$, and $m$ the
half-integer azimuthal angular momentum. We use the dimensionless
parameters $\alpha=M\mu$, $\chi=a/M$, and $\widehat\omega=M\omega$,
with $\widehat\omega_R=M\omega_R$ and $\widehat\omega_I=M\omega_I$.

\subsection{Master variables and angular spectrum}
\label{subsec:angular}
Let $F_{ab}=\nabla_a\psi_b-\nabla_b\psi_a$ be the field curl,
$C_b=\gamma^aF_{ab}$ its gamma contraction, and
$\Gamma_C=\gamma^bC_b$ its second contraction. The algebraic projector $\Pi$
onto gamma-traceless two-form--spinors is
\begin{equation}
 (\Pi F)_{ab}=F_{ab}
 -\frac12(\gamma_aC_b-\gamma_bC_a)
 +\frac16\gamma_{ab}\Gamma_C .
 \label{eq:curlprojector}
\end{equation}
Clifford algebra gives $\Pi^2=\Pi$,
$\gamma^a(\Pi F)_{ab}=0$, and complex image dimension eight.

The extraction subscripts denote spin weights, the rotation-generator
eigenvalues of the local angular frame. Stack the curl into a
24-component column $F$ with blocks
$(01,02,03,12,13,23)$. The field-master amplitudes $p_{\pm1/2}$
and their extraction matrix $P_F$ are defined by
\begin{equation}
 (p_{1/2},p_{-1/2})^{\mathsf T}=P_F\psi,\qquad
 P_F=\begin{pmatrix}
 1&1&1&1&-1&-1&-1&-1&0&0&0&0&0&0&0&0\\
 1&-1&-1&1&-1&1&1&-1&0&0&0&0&0&0&0&0
 \end{pmatrix}.
 \label{eq:PF}
\end{equation}
The curl-master amplitudes are
$(c_{-3/2},c_{3/2})^{\mathsf T}=P_EF$, where the extraction matrix is
\begin{equation}
 P_E=\frac1{16}
 \begin{pmatrix}0&A_E&B_E&-A_E&-B_E&0\end{pmatrix},
 \label{eq:PE}
\end{equation}
where each block has two rows and four columns, with
\begin{equation}
 A_E=\begin{pmatrix}
 \ii&-\ii&-\ii&\ii\\
 -\ii&-\ii&-\ii&-\ii
 \end{pmatrix},\qquad
 B_E=\begin{pmatrix}
 -1&1&1&-1\\
 -1&-1&-1&-1
 \end{pmatrix}.
\end{equation}
These rows include the projection \eqref{eq:curlprojector}.
Writing $\bm\Omega_{\rm old}=(p_{1/2},p_{-1/2},c_{-3/2},c_{3/2})^{\mathsf T}$,
define the normalized column
$\bm\Omega=(\Omega_{3/2},\Omega_{1/2},\Omega_{-1/2},\Omega_{-3/2})^{\mathsf T}$
by the constant matrix $T_\mu$:
\begin{equation}
 \bm\Omega_{\rm old}=T_\mu\bm\Omega,\qquad
 T_\mu=\begin{pmatrix}
 0&1&0&0\\0&0&-\ii&0\\
 0&0&0&-\ii\sqrt3\mu/16\\
 \sqrt3\mu/16&0&0&0
 \end{pmatrix}.
 \label{eq:Tmu}
\end{equation}

For the ordered spin weights $s=3/2,1/2,-1/2,-3/2$, the
spin-$3/2$ generators are
\begin{equation}
\begin{aligned}
 J_z&=\diag(3/2,1/2,-1/2,-3/2),\quad
 (J_+)_{i,i+1}=(\sqrt3,2,\sqrt3)_i,\quad J_-=J_+^\dagger,\\
 J_x&=(J_++J_-)/2,\quad J_y=(J_+-J_-)/(2\ii).
\end{aligned}
\end{equation}
Here $i=1,2,3$ and all other entries of $J_+$ vanish.
Define $q_\theta=m/W-a\omega W$. The matrix angular operators are
\begin{equation}
\begin{aligned}
 K_1={}&\frac W\mu J_y\partial_y-\frac{yJ_y}{2\mu W}
 +\frac{\ii q_\theta}{\mu}J_x-\frac{\ii y}{2\mu W}\{J_x,J_z\}
 -\ii ayJ_z+\frac{\ii\omega}{\mu^2}J_z^2,
\end{aligned}
 \label{eq:K1full}
\end{equation}
\begin{equation}
\begin{aligned}
 K_2={}&(1-y^2)\partial_y^2-2y\partial_y
 -\frac{m^2\id-2myJ_z+J_z^2}{1-y^2}
 +a^2(\omega^2-\mu^2)y^2\id+2a\omega yJ_z+2a\mu WJ_x .
\end{aligned}
 \label{eq:K2full}
\end{equation}
The derivative of $W$ is $-y/W$. Computing each derivative-order
coefficient gives $[K_1,K_2]=0$.

Let $\bm S=(S_{3/2},S_{1/2},S_{-1/2},S_{-3/2})^{\mathsf T}$ be a
regular joint angular profile satisfying
$K_1\bm S=\lambda\bm S$ and $K_2\bm S=\nu\bm S$.
The separation constants $\lambda,\nu$ enter the shifted variables
\begin{equation}
 x=\mu^2\lambda-\frac54\ii\omega,\qquad
 z=\nu-a^2\omega^2+2am\omega+\frac34.
 \label{eq:shifted}
\end{equation}
The spectral polynomial, distinct from the field curl $F_{ab}$, is
\begin{equation}
\begin{aligned}
 F(x,z)={}&16x^4+(16\mu^2+32\omega^2-40\mu^2z)x^2\\
 &+32\ii\mu^2[\omega(z-1)+3\mu^2(am-a^2\omega)]x
 +9\mu^4z^2-8\mu^2\omega^2z-16\mu^2\omega^2+16\omega^4\\
 &-36a^2\mu^6+96\mu^4(am\omega-a^2\omega^2).
\end{aligned}
 \label{eq:Ffull}
\end{equation}
Define the shifted angular operators
$X=\mu^2K_1-\frac54\ii\omega$ and
$Z=K_2-a^2\omega^2+2am\omega+\frac34$
so that $F(X,Z)=0$ is a differential-operator identity.
The main-text notation $F(\lambda,\nu)=0$ includes these shifts.

In the two regular spin patches the components obey
\begin{equation}
 S_s(y)=(1-y)^{|m-s|/2}(1+y)^{|m+s|/2}
       \times\text{a smooth function},\qquad
 s=3/2,1/2,-1/2,-3/2.
 \label{eq:angularends}
\end{equation}
For real parameters, $K_2$ is formally self-adjoint and elliptic.
Its rotation corrections are smooth zero-order bundle maps on the
sphere. Its regular realization has discrete spectrum.
The anti-self-adjoint $K_1$ preserves each finite-dimensional
$K_2$ eigenspace and is diagonalizable there, giving a joint complete
angular basis at real $\omega$.

At $a=0$, $j$ denotes total angular momentum and $\nu=-j(j+1)$;
components with $|s|>j$ are absent.
The $j=1/2$ sector is therefore treated in its allowed two-component
space. For complex $\omega$ we continue individual joint modes.
The real-frequency argument does not establish complex spectral
completeness \cite{Michishita2019}.

Numerically we use Jacobi basis functions $b_{n,s}$, with basis degree
$n=0,1,\ldots$ (distinct from the bound-state principal number),
Jacobi polynomial $P_n^{(\alpha_s,\beta_s)}$, and normalization
$\mathcal N_{n,s}$ in the measure $\dd y$. Here $\alpha_s,\beta_s$ are
the endpoint exponents and $\Gamma$ denotes Euler's gamma function:
\begin{equation}
\begin{aligned}
 b_{n,s}&=\mathcal N_{n,s}^{-1/2}
 (1-y)^{\alpha_s/2}(1+y)^{\beta_s/2}
 P_n^{(\alpha_s,\beta_s)}(y),\quad
 \alpha_s=|m-s|,\quad\beta_s=|m+s|,\\
 \mathcal N_{n,s}&=
 \frac{2^{\alpha_s+\beta_s+1}
 \Gamma(n+\alpha_s+1)\Gamma(n+\beta_s+1)}
 {(2n+\alpha_s+\beta_s+1)\Gamma(n+1)
 \Gamma(n+\alpha_s+\beta_s+1)} .
\end{aligned}
 \label{eq:jacobi}
\end{equation}
For complex frequency we use a non-Hermitian solve, continue branches
by eigenvector overlap, and check both angular residuals and the quartic.

\subsection{Radial transport and first integral}
\label{subsec:radial}
Applying the full RS equations and curl identities to the extraction
defines the master transport operator $\cP_3$:
\begin{equation}
 \cP_3\bm\Omega=
 (\partial_r+\ii\omega)[(\cB+K_2)\bm\Omega]
 +(2\mu^2K_1-3\mu^2r-\frac92\ii\omega)\bm\Omega=0,
 \label{eq:mastertransport}
\end{equation}
where the radial operator $\cB$ has scalar coefficients $B,V$:
\begin{equation}
 \cB=\Delta\partial_r^2+B\partial_r+V,\quad
 B=5(r-M)-4\ii Mr\omega+2\ii am,\quad
 V=(\omega^2-\mu^2)r^2+2M\omega^2r-3\ii\omega r
       -5\ii M\omega+\frac{15}{4}.
 \label{eq:radialB}
\end{equation}
The transport construction \cite{DiasSantos2026} closes after one radial
differentiation and initially admits one extra radial solution.

For a joint angular mode write
$\bm\Omega=e^{-\ii\omega T+\ii m\varphi}f(r)\bm S(\theta)$,
where $f$ is its scalar radial amplitude. The cubic adjoint polynomial
$A(r)$ and its constant term $A_0$ are
\begin{equation}
\begin{aligned}
 A(r)&=\mu^3r^3+2\mu(x-\ii\omega)r^2
 +\frac{4x^2+4\omega^2-3\mu^2z}{2\mu}\,r+A_0,
\end{aligned}
 \label{eq:A}
\end{equation}
\begin{equation}
\begin{aligned}
 A_0&=\frac1{3\mu^3}\big[
 6M\mu^4-12\ii a^2\mu^4\omega+12\ii am\mu^4
 +\ii\mu^2\omega z-4\ii\mu^2\omega-7\mu^2xz+4\mu^2x\\
 &\hspace{25mm}+4\ii\omega^3+4\omega^2x
 +4\ii\omega x^2+4x^3\big].
\end{aligned}
 \label{eq:A0}
\end{equation}
The physical second-order operator $L_2$, with primes denoting
$r$ derivatives, is
\begin{equation}
 L_2=A(\cB+\nu)-\Delta A'\partial_r
       +\Delta A''+(\Delta'-B+\ii\omega\Delta)A' .
 \label{eq:L2full}
\end{equation}
The radial transport operator is
$L_3=(\partial_r+\ii\omega)(\cB+\nu)
+2\mu^2\lambda-3\mu^2r-9\ii\omega/2$.
Coefficient comparison gives
\begin{equation}
 (\partial_r+\ii\omega)L_2-A L_3
       =-\frac{F(x,z)}{6\mu^3}\id .
 \label{eq:radialidentity}
\end{equation}
Thus $F=0$ permits $L_2f=C_{\rm int} e^{-\ii\omega r}$ under $L_3f=0$,
where $C_{\rm int}$ is an integration constant.
The extra relation $L_2f=0$ is obtained from the original equations in
the reconstruction elimination. It removes this integration constant;
it is not a freely imposed asymptotic condition.

Replacing the angular eigenvalues by $K_1,K_2$ in \eqref{eq:L2full}
defines the master first integral $\cP_5$. Its radial order is two;
the subscript is an operator label, not its radial derivative order.
The master system is $\cP_3\bm\Omega=\cP_5\bm\Omega=0$.
At a zero of $A$, this joint system supplies the continuation without
division by $A$.

The KS and BL radial amplitudes, denoted $f_{\rm KS}$ and $f_{\rm BL}$,
obey
\begin{equation}
 f_{\rm KS}=
 \exp\left[\ii\int^r\frac{2Mr'\omega-am}{\Delta(r')}\dd r'\right]
 f_{\rm BL}.
 \label{eq:radialphase}
\end{equation}
Frequencies and lifetimes are unchanged. The common
$2\ii M\omega\log r$ phase in the far KS amplitude is retained in all
matching calculations.

\subsection{Finite reconstruction and regularity}
\label{subsec:reconstruction}
Let $\cM\psi=\bm\Omega$ be the normalized master extraction, and let
$\cR$ map master solutions back to the field.
The reconstruction can be specified by a finite elimination from the
tetrad, Clifford matrices, and extraction above. The following
prescription fixes every coefficient without auxiliary files.

For a Lorentz generator $O^A{}_B$, with
$O_{AB}=\eta_{AC}O^C{}_B$, define its covector--spinor action
$\mathbb A(O)$ by
\begin{equation}
 \mathbb A(O)=\id_4\otimes
 \left(\frac14O_{AB}\gamma^{AB}\right)-O^{\mathsf T}\otimes\id_4 .
 \label{eq:Lorentzaction}
\end{equation}
Here $\otimes$ is the tensor product on the vector and spinor spaces.
Let $b$ generate a boost in the $0$--$1$ plane and $q$ a rotation in
the $2$--$3$ plane: $b^0{}_1=b^1{}_0=1$, $q^2{}_3=1$,
$q^3{}_2=-1$, with all other entries zero. Define the spatial-block
projection by $P\psi=(\psi_1,\psi_2,\psi_3)^{\mathsf T}$.
The commuting matrices $P\mathbb A(b)N$ and $P\mathbb A(q)N$ have
one-dimensional joint eigenspaces. The boost weight is the eigenvalue
of $P\mathbb A(b)N$. Form the constant matrix $S$ from their
eigenvectors, in descending boost weight and then descending spin
weight (the rotation eigenvalue divided by $\ii$), normalizing each
column's first nonzero entry to one. Their
ordered weights are
\begin{equation}
 \begin{array}{c|c}
 \text{boost weight}&\text{spin weights}\\ \hline
 3/2&1/2,\,-1/2\\
 1/2&3/2,\,1/2,\,-1/2,\,-3/2\\
 -1/2&3/2,\,1/2,\,-1/2,\,-3/2\\
 -3/2&1/2,\,-1/2
 \end{array}.
 \label{eq:weightbasis}
\end{equation}
Write the Fourier-mode field as
$\psi=NSv=e^{-\ii\omega r}NSU$, where $v$ and $U$ are twelve-component
columns before and after removing the radial phase. Define the
phase-shifted master column $H=e^{\ii\omega r}\bm\Omega_{\rm old}$.
All row and component indices below start at zero.

For a four-spinor row define
\begin{equation}
e_+=(1,1,1,1),\quad e_-=(1,-1,-1,1),\quad
 f_+=(1,1,-1,-1),\quad f_-=(1,-1,1,-1).
 \notag
\end{equation}
Compute the sixteen-component residual
\begin{equation}
X[U]=e^{\ii\omega r}(\slD+\mu)
          [e^{-\ii\omega r}NSU]
 \notag
\end{equation}
with $\partial_T=-\ii\omega$, $\partial_\varphi=\ii m$, using
the full covector and spin connections in \eqref{eq:covder}.
From its four spinor blocks $X_A$, form the residual rows
\begin{equation}
\begin{aligned}
 (D_0,\ldots,D_{15})^{\mathsf T}
 =\big(&e_+(X_0-X_1),\,e_-(X_0-X_1),\,
 e_+(X_2-\ii X_3),\,f_+(X_0-X_1),\\
 &e_-(X_2-\ii X_3),\,f_-(X_0-X_1),\,
 e_+(X_2+\ii X_3),\,e_-(X_2+\ii X_3),\\
 &f_+(X_2-\ii X_3),\,e_+(X_0+X_1),\,
 f_-(X_2-\ii X_3),\,e_-(X_0+X_1),\\
 &f_+(X_2+\ii X_3),\,f_-(X_2+\ii X_3),\,
 f_+(X_0+X_1),\,f_-(X_0+X_1)\big)^{\mathsf T}.
\end{aligned}
 \label{eq:residualrows}
\end{equation}
Likewise let $C=e^{\ii\omega r}\cM_{\rm old}
[e^{-\ii\omega r}NSU]$, where $\cM_{\rm old}$ is the field/curl
extraction in \eqref{eq:PF} and \eqref{eq:PE}.
The entries $H_i,C_i$, $i=0,1,2,3$, follow the old master ordering.

Introduce four auxiliary scalar functions $h_j(r,\theta)$,
$j=2,3,4,5$, and start with
\begin{equation}
 U_0=-H_0/8,\quad U_1=-H_1/8,\quad
 (U_2,U_3,U_4,U_5)=
 (h_2/z_+,\,h_3/z_+^2,\,h_4/z_-^2,\,h_5/z_-),
 \quad z_\pm=r\pm\ii ay,
 \label{eq:initialreconstruction}
\end{equation}
and $U_6=\cdots=U_{11}=0$. The four residuals
$D_0,D_1,C_3-H_3,C_2-H_2$ determine
\begin{equation}
 \begin{array}{c|c}
 \text{transport}&\text{residual evaluated with all }\partial_rh_j=0\\ \hline
 \partial_rh_3&(\ii z_+^2/8)D_0\\
 \partial_rh_4&(\ii z_-^2/8)D_1\\
 \partial_rh_2&2\ii z_+(C_3-H_3)\\
 \partial_rh_5&-2\ii z_-(C_2-H_2)
 \end{array}
 \quad\equiv G_jH .
 \label{eq:hsource}
\end{equation}
The row differential operators $G_j$ are defined by the table's residual
formulas; their right sides contain only $H$ and its derivatives. Use these
transports, including their angular derivatives, in every subsequent
radial differentiation. Recover the other six components sequentially;
the table specifies the residual row $i$ and scalar multiplier $c_k$
for each target component $U_k$:
\begin{equation}
 \begin{array}{c|rrrrrr}
 \text{target }U_k&U_7&U_8&U_6&U_9&U_{10}&U_{11}\\ \hline
 \text{residual }D_i&D_3&D_5&D_8&D_{13}&D_{10}&D_{12}\\
 \text{multiplier }c_k&
 -\ii z_+/8&-\ii z_-/8&-1/(8\mu)&-1/(8\mu)&z_-/8&z_+/8
 \end{array},
 \qquad U_k=c_kD_i\big|_{U_k=0}.
 \label{eq:lowerreconstruction}
\end{equation}
Unassigned components remain zero at each step.

For the remaining residuals, put $t_\theta=\tan(\theta/2)$ and
$\Sigma_p=(1+t_\theta^2)^2\Sigma$.
Clear the denominators of the coefficients multiplying $h_j$ and
their angular derivatives by the least required powers of
$t_\theta,1+t_\theta^2,\Sigma_p,\mu$.
Call the resulting rows $E_i$, without further rescaling. In the next
two equations, $J_i[H]$ denotes the part containing only the masters
after the indicated differentiation and substitutions; $d$ and
$b_{\rm alg}$ are the scalar coefficients defined there.
Using \eqref{eq:hsource}, five radial derivatives of $E_2,E_7$ give
\begin{equation}
 \partial_r^5 E_2=d\,h_2+J_2[H],\qquad
 \partial_r^5 E_7=d\,h_5+J_7[H],\qquad
 d=960\mu^2t_\theta^2(1+t_\theta^2)^6 .
 \label{eq:polynomialextraction}
\end{equation}
Thus $h_2=-J_2/d$, $h_5=-J_7/d$. Substituting these expressions into
$E_4,E_6$ leaves
\begin{equation}
 E_4=b_{\rm alg}h_3+J_4[H],\qquad E_6=b_{\rm alg}h_4+J_6[H],\qquad
 b_{\rm alg}=16\ii\mu t_\theta(1+t_\theta^2)^2 ,
 \label{eq:middleextraction}
\end{equation}
Thus $h_3=-J_4/b_{\rm alg}$ and $h_4=-J_6/b_{\rm alg}$.
Equations~\eqref{eq:initialreconstruction}--\eqref{eq:middleextraction}
are a closed finite differential construction of $\cR$.

The hatted master equations are obtained by the change $T_\mu$ and
radial phase conjugation defining $H$. Reduction by these equations,
using leading derivative
monomials $\partial_r^3$ and
$\partial_r^2\partial_{t_\theta}^3$, gives $h_j$ of orders
$(2,3,3,2)$ and an order-four field representative at fixed
$(\omega,m)$. For example,
\begin{equation}
 h_2=-\frac{2\ii}{3\mu^2}
 (\widehat{\cB}+K_{2,+3/2,+3/2})H_3,\qquad
 h_5=\frac{2\ii}{3\mu^2}
 (\widehat{\cB}+K_{2,-3/2,-3/2})H_2,\qquad
 \widehat{\cB}=e^{\ii\omega r}\cB e^{-\ii\omega r}.
 \label{eq:hcompact}
\end{equation}
The subscripts on $K_2$ denote diagonal entries in the normalized
spin-weight basis; $H_2,H_3$ retain the old curl normalization.

Let $\mathcal E$ denote the column of the sixteen RS field equations.
Substitution of the construction into the original equations and
re-extraction, with reduction by the two master equations, gives
\begin{equation}
 \mathcal E\cR\bm\Omega=0,\qquad
 \cM\cR\bm\Omega=\bm\Omega .
 \label{eq:converse}
\end{equation}
To prove injectivity, set $H=0$. The transports make $h_j$
independent of $r$; \eqref{eq:polynomialextraction} then forces
$h_2=h_5=0$, and \eqref{eq:middleextraction} forces $h_3=h_4=0$
for $\mu\ne0$. All twelve components vanish.
Applying this to $\psi-\cR\cM\psi$ proves the other inverse identity.

The reduced coefficients have denominators only in
$t_\theta,1+t_\theta^2,\Sigma_p,\mu$.
A smooth representative is obtained by using the reduced $h_j$ and
retaining the last two field formulas in \eqref{eq:lowerreconstruction}
before their final master reduction. Its total spacetime differential
order is six. Near each axis, transform spin-weight $s$ inputs by
$e^{-\ii s\varphi}$ in the north patch and $e^{+\ii s\varphi}$ in the
south patch. Use the local complex coordinate
$\zeta=t_\theta e^{\ii\varphi}$ at the north pole and
$\zeta=t_\theta^{-1}e^{\ii\varphi}$ at the south pole, with
$\bar\zeta$ its complex conjugate. For nonnegative integers $K,L$, let
$(e_\zeta,e_{\bar\zeta})$ be the integer shifts of the powers of
$\zeta,\bar\zeta$ when a block acts on $\zeta^K\bar\zeta^L$.
Each such coefficient, after clearing
smooth nonzero denominators, is divisible by
\begin{equation}
 (K)_{\max(0,-e_\zeta)}(L)_{\max(0,-e_{\bar\zeta})}.
 \label{eq:polartest}
\end{equation}
Here $(K)_p=K(K-1)\cdots(K-p+1)$ is the falling factorial, with
$(K)_0=1$. These identities give smooth Cartesian differential
coefficients at both poles. There are no inverse $\Delta$,
ergosurface, or spectral factors. The off-horizon identities therefore
extend to the future horizon by smoothness, including extremal
parameters. This continuation does not solve the extremal spectral
boundary problem.

\section{Quasibound spectra and rotational splitting}
\label{sec:bound}
\subsection{Endpoint conditions}
Write $L_2f=q_2f''+q_1f'+q_0f=0$, where $q_0,q_1,q_2$ are the scalar
radial coefficient polynomials
\begin{equation}
\begin{aligned}
 q_2&=A\Delta,\qquad q_1=AB-\Delta A',\\
 q_0&=A(V+\nu)+\Delta A''+(\Delta'-B+\ii\omega\Delta)A'.
\end{aligned}
 \label{eq:qcoeff}
\end{equation}
At a subextremal future horizon, let $f_H$ denote the ingoing radial
solution. It has a regular series $f_H=\sum_{n=0}^\infty c_n(r-r_+)^n$,
where $c_n$ are Frobenius coefficients. Define the Taylor coefficients
$q_{k,j}$ by $q_k=\sum_{j\ge0}q_{k,j}(r-r_+)^j$, $k=0,1,2$;
then $q_{2,0}=0$. In the recursion below, $n,j$ are nonnegative series
indices, distinct from the spectral labels introduced later:
\begin{equation}
\begin{aligned}
 &(n+1)(q_{1,0}+nq_{2,1})c_{n+1}
 =-\sum_{j=0}^{n}q_{0,j}c_{n-j}
 -\sum_{j=1}^{n}q_{1,j}(n-j+1)c_{n-j+1}\\
 &\hspace{43mm}
 -\sum_{j=2}^{n+1}q_{2,j}(n-j+2)(n-j+1)c_{n-j+2}.
\end{aligned}
 \label{eq:horizonrec}
\end{equation}
Terms outside the polynomial degrees vanish. Exceptional recursion
denominators require the joint system; the reported roots do not use
division at such an exception.

At infinity, $q$ is the inverse decay length, $f_\infty$ the decaying
solution, $p_\infty$ its Coulomb power, and $N_\infty$ the truncation
order of its inverse-radius series with coefficients $d_n$:
\begin{equation}
 q=\sqrt{\mu^2-\omega^2},\quad\operatorname{Re}q>0,\qquad
 f_\infty=e^{-qr}r^{p_\infty}\sum_{n=0}^{N_\infty}d_n r^{-n},\qquad
 p_\infty=-1+2\ii M\omega+\frac{M(2\omega^2-\mu^2)}q .
 \label{eq:boundinfinity}
\end{equation}
The coefficients follow by substitution into \eqref{eq:qcoeff}.
The Coulomb power is retained at weak binding.

At each complex trial frequency the angular problem is solved again.
Integrating the two endpoint solutions to the matching radius $r_{\rm m}$ gives
\begin{equation}
 \frac{f_H'}{f_H}(r_{\rm m})-
 \frac{f_\infty'}{f_\infty}(r_{\rm m})=0.
 \label{eq:boundmatch}
\end{equation}
We avoid nodes, check the normalized Wronskian, and polish small decay
rates with a local complex Newton correction.

\subsection{Orbital labels and mode families}
Let $\ell$ be orbital angular momentum, $n_r\ge0$ the radial overtone,
and $n=n_r+\ell+1$ the principal number in the weak-binding limit.
At $M=1$, define the scaled radius $\rho=\mu^2r$, the binding-energy
coefficient $e$ by $\omega=\mu+\mu^3e$, and the leading angular
coefficient $\kappa_1$ by $\mu^2\lambda=\ii\mu\kappa_1+O(\mu^3)$.
The leading exterior equation is
\begin{equation}
 f_{\rho\rho}+\frac2\rho f_\rho+
 \left(2e+\frac2\rho-\frac{\ell(\ell+1)}{\rho^2}\right)f=0,\qquad
 \ell(\ell+1)=j(j+1)-2\kappa_1+\frac{15}{4}.
 \label{eq:hydrogen}
\end{equation}
Hence $e=-1/(2n^2)$, $n=n_r+\ell+1$.
For the lowest angular sectors,
\begin{equation}
 \begin{array}{c|c|c}
 j&\kappa_1&\ell\\\hline
 1/2&-3/4,\ 5/4&2,\ 1\\
 3/2&-9/4,\ 3/4,\ 11/4,\ 15/4&3,\ 2,\ 1,\ 0
 \end{array}.
 \label{eq:orbitaltable}
\end{equation}
For Dirac, $\ell=j\pm1/2$ \cite{DolanDempsey2015}.
The labels here are fixed by continuation to weak binding.

With time dependence $e^{-\ii\omega T}$, a mode with $\omega_I<0$ has
amplitude decay time $\tau_{\rm amp}$ and occupation decay time $\tau_N$:
\begin{equation}
 \tau_{\rm amp}=-1/\omega_I,\qquad \tau_N=-1/(2\omega_I).
 \label{eq:lifetime}
\end{equation}
The latter is the occupation decay time of a linear mode.

At $\alpha=0.2$ the Schwarzschild starting points are
\begin{equation}
 \begin{array}{c|c|c|c}
 \text{family}&(n,\ell,j)&M\omega_R&M\omega_I\\\hline
 A&(3,2,1/2)&0.199541234024125&-6.1482950\,10^{-10}\\
 B&(2,1,1/2)&0.198835889957086&-1.66199728\,10^{-5}\\
 C&(1,0,3/2)&0.195749060718686&-7.06080806\,10^{-6}\\
 D&(2,1,3/2)&0.198948223738201&-2.08450267\,10^{-8}
 \end{array}.
 \label{eq:frequencytable}
\end{equation}
Families A--D are followed with $m=\pm j$ at
\begin{equation}
 \chi\in\{0,\ 0.15,\ 0.3,\ 0.45,\ 0.6,\ 0.75,\ 0.85,\ 0.9,\ 0.95\}.
 \notag
\end{equation}

\begin{figure}[tbp]
 \centering
 \makebox[\columnwidth][c]{\hspace*{-0.22in}\includegraphics[width=3.375in]{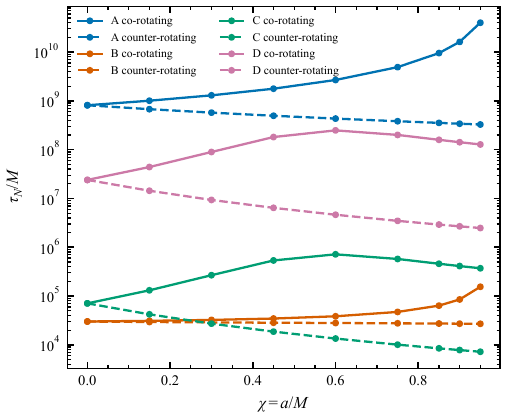}\hspace*{0.22in}}
 \caption{Selected quasibound occupation decay times at $\alpha=0.2$.
 Solid and dashed curves denote $m=+j$ and $m=-j$.
 All plotted modes decay.}
 \label{fig:lifetimes}
\end{figure}

Fig.~\ref{fig:lifetimes} shows the lifetimes of these families for
$m=\pm j$; selected values appear in Table~\ref{tab:lifetime}.

\begin{table}[tbp]
 \caption{Occupation decay times at $\alpha=0.2$, in units of $M$.
 The amplitude time is twice the listed value.}
 \begin{ruledtabular}
 \begin{tabular}{c r r r}
 Family&$\chi=0$&$\chi=0.9,\ m=+j$&$\chi=0.9,\ m=-j$\\
 A&$8.13\,10^8$&$1.62\,10^{10}$&$3.40\,10^8$\\
 B&$3.01\,10^4$&$8.52\,10^4$&$2.72\,10^4$\\
 C&$7.08\,10^4$&$4.12\,10^5$&$7.81\,10^3$\\
 D&$2.40\,10^7$&$1.42\,10^8$&$2.68\,10^6$
 \end{tabular}
 \end{ruledtabular}
 \label{tab:lifetime}
\end{table}
The C and D co-rotating families have local lifetime maxima near
$\chi=0.5947,0.5996$, with
$\tau_N/M\simeq7.17\,10^5,2.48\,10^8$.
A matched Dirac calculation also has lifetime extrema.
These sampled families do not establish stability of all modes.

\subsection{Rotational splitting and the Dirac comparison}
\label{subsec:splitting}
Select $(n,\ell,j)=(2,1,1/2)$, the B family. Set $M=1$ and use the
scalar mass abbreviation $u=\mu$ within this subsection.
Denote the zero-rotation angular eigenvalues by $\lambda_0,\nu_0$.
They and their derivatives at fixed $\omega$ are
\begin{equation}
 \lambda_0=\ii\left(\frac{\omega}{4u^2}+\frac1u\right),\quad
 \nu_0=-\frac34,\qquad
 \left.\partial_a\lambda\right|_0=
 -\frac{\ii m}{3}\left(1+\frac{4\omega}{u}\right),\quad
 \left.\partial_a\nu\right|_0=\frac{2m}{3}(\omega+4u).
 \label{eq:angularpert}
\end{equation}
For $m=1/2$, the allowed profiles are
$\sqrt{(1+y)/2}$ and $\sqrt{(1-y)/2}$. Their diagonal $y$ expectation
and off-diagonal $W$ overlap are $1/3$ and $2/3$.
First-order perturbation of both commuting operators gives
\eqref{eq:angularpert}; reflection gives negative $m$.

Introduce the rotational frequency coefficient $c$ by
\begin{equation}
 \rho=u^2r,\qquad \omega=u+u^3e+a u^6c .
 \label{eq:scaling}
\end{equation}
After division by $q_2$ and conversion to $\rho$, define the scalar
coefficients $P_\rho,Q_\rho$ by
$f_{\rho\rho}+P_\rho f_\rho+Q_\rho f=0$.
The reduced wavefunction $U$ and normal-form potential $\mathcal V$ satisfy
\begin{equation}
 f=e^{-\frac12\int P_\rho\,\dd\rho}U,\qquad
 U_{\rho\rho}+\mathcal VU=0,\qquad
 \mathcal V=Q_\rho-\frac12\partial_\rho P_\rho-\frac14P_\rho^2.
 \label{eq:normalform}
\end{equation}
Removing the first derivative eliminates lower-order imaginary radial
phase terms. Denote the coefficient of $a u^3$ in this potential by
$\mathcal V^{\rm RS}_{a u^3}$:
\begin{equation}
 \mathcal V^{\rm RS}_{a u^3}
 =2c-\frac{16em}{3\rho^2}
      -\frac{4m}{\rho^3}+\frac{20m}{3\rho^4}.
 \label{eq:VRS}
\end{equation}
For $n=2,\ell=1$, the unperturbed solution is
$U_0=\rho^2e^{-\rho/2}$ with $e=-1/8$.
The first-order solvability condition gives
\begin{equation}
\begin{aligned}
 0&=\int_0^\infty U_0^2\mathcal V^{\rm RS}_{a u^3}\,\dd\rho
 =48c-\frac{32em}{3}-4m+\frac{20m}{3}
 =48c+4m ,
\end{aligned}
 \label{eq:solvabilityRS}
\end{equation}
Thus the RS rotational coefficient is $c_{\rm RS}=-m/12$.
This leading exterior calculation does not give a uniform remainder
bound for the horizon boundary layer.

For the Dirac comparison, $R_1,R_2$ are BL radial amplitudes,
$\lambda_D$ their angular separation constant, and
$K=(r^2+a^2)\omega-am$ the radial frequency combination.
The published equations are \cite{DolanDempsey2015}
\begin{equation}
\begin{aligned}
 \sqrt\Delta(\partial_r-\ii K/\Delta)R_1
       &=(\lambda_D+\ii\mu r)R_2,\\
 \sqrt\Delta(\partial_r+\ii K/\Delta)R_2
       &=(\lambda_D-\ii\mu r)R_1,\qquad K=(r^2+a^2)\omega-am .
\end{aligned}
 \label{eq:Diracpair}
\end{equation}
For $j=1/2,\ell=1$, $\lambda_D=+1$ at $a=0$, and
\begin{equation}
 \left.\partial_a\lambda_D\right|_0=\frac{2m}{3}(u-2\omega).
\end{equation}
Factor $R_2=e^{-\ii\int K\,\dd r/\Delta}f$, eliminate $R_1$, and
make the same normal-form and weak-binding reduction. The Dirac
coefficient of $au^3$ in the potential is
\begin{equation}
 \mathcal V^{\rm D}_{a u^3}
 =2c+\frac{8em}{3\rho^2}
      -\frac{4m}{\rho^3}-\frac{10m}{3\rho^4},
\end{equation}
and therefore
\begin{equation}
 0=48c+\frac{16em}{3}-4m-\frac{10m}{3}
   =48c-8m,\qquad c_D=\frac m6.
 \label{eq:solvabilityD}
\end{equation}
The comparison matches orbital, principal, and total angular momentum
labels as well as the frequency normalization.

Let $\bm L$ denote orbital angular momentum, $\bm{\mathsf S}$ intrinsic
spin, and $\bm n$ the radial unit vector, with the $z$ axis along the
black-hole spin. The leading weak-field rotational Hamiltonian is
\cite{Obukhov2009}
\begin{equation}
 H_J=\frac{aM}{r^3}
 [2L_z+3n_z(\bm n\cdot\bm{\mathsf S})-\mathsf S_z].
 \label{eq:LT}
\end{equation}
For $\ell=1,j=m=1/2$, let $s_{\rm int}$ denote intrinsic spin and
$\langle\cdot\rangle$ the expectation value in that angular state.
The matrix elements are
\begin{equation}
 \begin{array}{c|r|r|r|r}
 s_{\rm int}&\langle L_z\rangle&\langle\mathsf S_z\rangle&
 \langle3n_z(\bm n\cdot\bm{\mathsf S})-\mathsf S_z\rangle&
 \text{total bracket}\\\hline
 1/2&2/3&-1/6&2/3&2\\
 3/2&-1/3&5/6&-1/3&-1
 \end{array}.
\end{equation}
Together with $\langle r^{-3}\rangle=u^6/24$ at $M=1$, this reproduces
the two coefficients. The field-equation calculation is the direct
derivation; the angular expectation explains the sign through familiar
frame dragging and angular-momentum addition.

Define $\Delta\widehat\omega_R=\widehat\omega_R(+1/2)-\widehat\omega_R(-1/2)$.
Consequently,
\begin{equation}
 \lim_{\alpha\to0}\alpha^{-6}
 \left.\partial_\chi\Delta\widehat\omega_R\right|_{\chi=0}
 =-\frac1{12}\ \text{(RS)},\qquad
 =+\frac16\ \text{(Dirac)} .
 \label{eq:limitstatement}
\end{equation}
The numerical scaled splittings are
\begin{equation}
 \begin{array}{c|r|r}
 \alpha&
 \Delta\widehat\omega_R/(\chi\alpha^6)\ \text{(RS)}&
 \Delta\widehat\omega_R/(\chi\alpha^6)\ \text{(Dirac)}\\\hline
 0.20&-0.23904&+0.27053\\
 0.10&-0.08256&+0.18540\\
 0.05&-0.08259&+0.17094\\
 0.035&-0.08296&+0.16873
 \end{array}.
\end{equation}
Both $\chi=0.01$ and $0.02$ are checked.
Simultaneously reversing $(a,m)$ preserves the frequency within
$3\,10^{-17}$ in these tests.
At the smallest mass the splitting is a difference of nearby
double-precision frequencies; its final digits are not certified.

\section{Scattering and Hawking emission}
\label{sec:flux}
The scattering calculation uses the same angular eigenfunctions, radial
equation, and field reconstruction as the bound-state calculation.
The complex frequencies of main-text Fig.~1 are roots of the
bound-state problem in Sec.~S2. Here the continuum frequency
$\omega>\mu$ is a real input. At each
$(\omega,m)$, the outputs are the reflection matrix, greybody factors,
and outgoing helicity content. These determine the particle spectra in
main-text Fig.~2.

\subsection{Physical flux and scattering normalization}
Let $b=1,\ldots,d_m$ label the $d_m$ retained joint angular modes at fixed
$m$. Denote their profiles and separation constants by
$\bm S_{mb}$, $\lambda_{mb}$, and $\nu_{mb}$, so that
$K_1\bm S_{mb}=\lambda_{mb}\bm S_{mb}$ and
$K_2\bm S_{mb}=\nu_{mb}\bm S_{mb}$. At real frequency,
$\lambda_{mb}$ is imaginary and $\nu_{mb}$ is real. These separation
constants are distinct from the physical helicity $h$ below.
For each branch, insert its separation constants into the radial
operator $L_2$ of Sec.~S1.3, whose coefficients are expanded in
\eqref{eq:qcoeff}. The radial solution
$f_{H,mb}$ is regular at the future horizon; its two infinity basis
solutions are $f_{-,mb}$ (incoming) and $f_{+,mb}$ (outgoing).
The corresponding physical fields are
\begin{equation}
 \Psi_{x,mb}
 =\cR\left[e^{-\ii\omega T+\ii m\varphi}
             f_{x,mb}(r)\bm S_{mb}(\theta)\right],
 \qquad x=H,-,+.
 \label{eq:scatteringfields}
\end{equation}
Thus the scalar amplitudes $f_{x,mb}$ are used to integrate and match
the radial equation, while the reconstructed covector--spinors
$\Psi_{x,mb}$ are used to evaluate physical currents.

For two solutions $\psi_1,\psi_2$, the action gives the conserved
sesquilinear current
\begin{equation}
 j^a[\psi_1,\psi_2]=-\ii\bar\psi_{1b}\gamma^{bac}\psi_{2c}.
 \label{eq:current}
\end{equation}
The frame current is converted to a coordinate current by
$j^r=E_a{}^rj^a$. At equal real $\omega,m$, define the outward flux pairing
\begin{equation}
 \Phi[\psi_1,\psi_2]
 =2\pi\int_{-1}^1\Sigma j^r[\psi_1,\psi_2]\,\dd y .
 \label{eq:fluxpairing}
\end{equation}
It is independent of $r$; write $\Phi[\psi]=\Phi[\psi,\psi]$.
On gamma-traceless fields the current reduces to
$j^a[\psi_1,\psi_2]=-\ii\bar\psi_{1b}\gamma^a\psi_2{}^b$
\cite{Schenkel2012}. The positive-frequency physical asymptotic modes have positive
density and radial velocity $\pm k/\omega$, where
$k=\sqrt{\omega^2-\mu^2}>0$ is the asymptotic momentum magnitude.

Suppressing $m,b$ temporarily, the infinity basis amplitudes are
\begin{equation}
 f_\pm=e^{\pm\ii kr}
 r^{-1+2\ii M\omega\pm\ii M(2\omega^2-\mu^2)/k}
       \sum_{n=0}^{N_\infty}c_n^\pm r^{-n},
 \qquad c_0^\pm=1.
 \label{eq:scatteringfar}
\end{equation}
Here $c_n^\pm$ are inverse-radius series coefficients and $N_\infty$
is their truncation order. Matching the horizon solution at a radius
$r_{\rm m}$ gives the coefficients $C_-,C_+$ through
\begin{equation}
 \begin{pmatrix}
 f_-(r_{\rm m})&f_+(r_{\rm m})\\
 f_-'(r_{\rm m})&f_+'(r_{\rm m})
 \end{pmatrix}
 \begin{pmatrix}C_-\\C_+\end{pmatrix}
 =
 \begin{pmatrix}f_H(r_{\rm m})\\f_H'(r_{\rm m})\end{pmatrix}.
 \label{eq:scatteringmatch}
\end{equation}
Primes denote $r$ derivatives. Define the positive infinity flux
norms $N_-=-\Phi[\Psi_-]$, $N_+=\Phi[\Psi_+]$ and the outward horizon
flux $\Phi_H=\Phi[\Psi_H]$. The pure incoming and outgoing infinity
subspaces are flux-orthogonal. For one channel the reflection and
absorption probabilities are
\begin{equation}
 \mathcal R_{\rm refl}=\frac{N_+|C_+|^2}{N_-|C_-|^2},
 \qquad
 \Gamma=-\frac{\Phi_H}{N_-|C_-|^2}=1-\mathcal R_{\rm refl}.
 \label{eq:singleGamma}
\end{equation}
The incident flux magnitude is
$\Phi_{\rm in}=N_-|C_-|^2$, as used in the main text.
Neither $|C_+|^2/|C_-|^2$ nor the squared master amplitude alone is a
physical probability; the normalization depends on the reconstructed
field. The overall normalization of $f_H$ cancels.

Restore the branch indices on $N_{\pm,mb}$, the flux norms defined
above. For several branches, first divide each infinity mode by its own norm:
\begin{equation}
 \widehat\Psi_{-,mb}=\frac{\Psi_{-,mb}}{\sqrt{N_{-,mb}}},
 \qquad
 \widehat\Psi_{+,mb}=\frac{\Psi_{+,mb}}{\sqrt{N_{+,mb}}},
 \qquad
 D_\pm=\operatorname{diag}_b\sqrt{N_{\pm,mb}} .
 \label{eq:individualfluxnorm}
\end{equation}
The $d_m\times d_m$ positive Gram matrices of these rescaled modes,
with row and column labels $b,c=1,\ldots,d_m$, are
\begin{equation}
 (G_{\rm in})_{bc}=-\Phi[\widehat\Psi_{-,mb},\widehat\Psi_{-,mc}],
 \qquad
 (G_{\rm out})_{bc}=\Phi[\widehat\Psi_{+,mb},\widehat\Psi_{+,mc}] .
 \label{eq:fluxgrams}
\end{equation}
Their diagonal entries are one; off-diagonal pairings are retained.
The reflected coefficient map in these bases is
\begin{equation}
 C_{\rm ref}
 =D_+\operatorname{diag}_b\left(\frac{C_{+,mb}}{C_{-,mb}}\right)D_-^{-1}.
 \label{eq:reflectionmap}
\end{equation}
Here $\operatorname{diag}_b$ forms a diagonal matrix in branch index
$b$; both $D_\pm$ are diagonal, positive matrices.
If $\bm c_{\rm in},\bm c_{\rm out}$ are coefficients in the rescaled
bases, their unit-flux coordinates are
$\bm a_{\rm in}=G_{\rm in}^{1/2}\bm c_{\rm in}$ and
$\bm a_{\rm out}=G_{\rm out}^{1/2}\bm c_{\rm out}$.
Positive Hermitian square roots therefore give
\begin{equation}
 \begin{aligned}
 \bm a_{\rm out}&=S_m\bm a_{\rm in},
 &S_m&=G_{\rm out}^{1/2}C_{\rm ref}G_{\rm in}^{-1/2},\\
 A_{{\rm abs},m}&=\id-S_m^\dagger S_m,
 &A_{{\rm em},m}&=\id-S_mS_m^\dagger .
 \end{aligned}
 \label{eq:Smatrix}
\end{equation}
This $S_m$ is the reflection matrix in the main text.
$A_{{\rm abs},m}$ acts on incident channels and
$A_{{\rm em},m}$ on outgoing channels. They have the same eigenvalues
$1-\sigma_b^2$, where $\sigma_b$ are the singular values of $S_m$,
and the same trace.

The branch curves in Fig.~\ref{fig:greybody} use
$\Gamma_{mb}=(A_{{\rm abs},m})_{bb}$ in this incoming unit-flux basis.
For a single flux-orthogonal branch this is
\eqref{eq:singleGamma}; with several branches the full matrices in
\eqref{eq:Smatrix} are used. The rank $b$ orders $\nu_{mb}$ from largest
to smallest, with $-\ii K_1$ resolving spherical degeneracies. This
angular label is distinct from helicity, which is determined from the
outgoing physical field below.

\subsection{Asymptotic helicity}
In the radial asymptotic frame, let $J_{23}^{(v)}$ generate rotations
of the covector index in the transverse $2$--$3$ plane. The total radial
spin generator $\mathsf H$ is
\begin{equation}
 \mathsf H=J_{23}^{(v)}\otimes\id+
      \id\otimes(-\ii\gamma^{23}/2),\qquad
 (J_{23}^{(v)})_{23}=-\ii,\quad (J_{23}^{(v)})_{32}=\ii .
\end{equation}
All other entries of $J_{23}^{(v)}$ vanish. The eigenvalues $h$ of
$\mathsf H$ are helicities. The longitudinal and transverse projectors are
\begin{equation}
 P_L=(9\id/4-\mathsf H^2)/2,\qquad P_T=\id-P_L .
 \label{eq:helicityprojector}
\end{equation}
They select $|h|=1/2$ and $3/2$ on the flat physical kernel, respectively,
and preserve its gamma-trace and divergence constraints.
Each sector contains two physical states.

To connect these projectors to the separated modes, expand the
reconstruction in \eqref{eq:scatteringfields} at large $r$ using
\eqref{eq:scatteringfar}. For each individually normalized outgoing
mode define the leading sixteen-component angular field $U_{mb}$ by
\begin{equation}
 \widehat\Psi_{+,mb}
 =e^{-\ii\omega T+\ii m\varphi+\ii kr}
       r^{\ii\eta_\infty-1}[U_{mb}(\theta)+O(r^{-1})],
 \qquad
 \eta_\infty=2M\omega+\frac{M(2\omega^2-\mu^2)}{k}.
 \label{eq:outgoingfield}
\end{equation}
The leading field follows from the same $\bm S_{mb}$ and
$f_{+,mb}$ used for scattering. The projection acts on $U_{mb}$,
after cancellation of positive radial powers in the reconstruction.

For angular fields $U,V$ in this outgoing space, the infinity flux is
\begin{equation}
 \mathcal F_\infty[U,V]
 =2\pi\int_{-1}^1U^\dagger\mathsf J_rV\,\dd y,
 \qquad
 \mathsf J_r=\operatorname{diag}(-1,1,1,1)
                 \otimes(-\ii\beta\gamma^1).
 \label{eq:infinitypairing}
\end{equation}
The covector blocks are ordered as in Sec.~S1.1, and
$\beta=-\ii\gamma^0$. In particular,
$(G_{\rm out})_{bc}=\mathcal F_\infty[U_{mb},U_{mc}]$.
Define the projected Gram matrix and its unit-flux representation by
\begin{equation}
 (G_L)_{bc}=\mathcal F_\infty[P_LU_{mb},P_LU_{mc}],
 \qquad
 \mathsf L_m=G_{\rm out}^{-1/2}G_LG_{\rm out}^{-1/2}.
 \label{eq:longitudinalmatrix}
\end{equation}
Thus $P_L$ acts on the sixteen physical field components, while
$\mathsf L_m$ acts on the $d_m$ outgoing scattering coefficients.
For an outgoing unit-flux vector $\bm a_{\rm out}$, its longitudinal
flux is $\bm a_{\rm out}^\dagger\mathsf L_m\bm a_{\rm out}$.
At finite angular cutoff, $\mathsf L_m$ is a compression of the full
projector, so its eigenvalues need not be exactly zero or one.
A separated angular mode generally contains both helicity sectors.

\subsection{Emission spectra and massless references}
\label{subsec:emission}
For the usual free-fermion Unruh prescription, the occupation at real
frequency $\omega$ and azimuthal number $m$ is
\begin{equation}
 n_F(\omega,m)=\frac1{e^{(\omega-m\Omega_H)/T_H}+1},\qquad
 \Omega_H=\frac{a}{2Mr_+},\qquad
 T_H=\frac{\sqrt{M^2-a^2}}{4\pi Mr_+}.
 \label{eq:occupation}
\end{equation}
Here $\Omega_H$ is the horizon angular velocity and $T_H$ the Hawking
temperature \cite{Hawking1975,Page1976}. All channels at the same
$(\omega,m)$ have the same occupation factor. Their outgoing spectral
occupation matrix is
\begin{equation}
 \mathsf W_m(\omega)=\frac{n_F(\omega,m)}{2\pi}A_{{\rm em},m}.
 \label{eq:emissionmatrix}
\end{equation}
Its trace gives the number emitted per unit time and frequency in that
$m$ sector; its contraction with $\mathsf L_m$ gives the longitudinal
part. Let $N$ be the emitted particle number and $\Tr$ the trace over
angular channels at fixed $m$. Denote the total, longitudinal, and
transverse number spectra by $Q,Q_L,Q_T$, respectively:
\begin{equation}
 \begin{aligned}
 Q(\omega)\equiv\frac{d^2N}{dt\,d\omega}
   &=\sum_m\Tr\mathsf W_m
    =\frac1{2\pi}\sum_m n_F\Tr A_{{\rm em},m},\\
 Q_L(\omega)&=\sum_m\Tr(\mathsf L_m\mathsf W_m)
    =\frac1{2\pi}\sum_m n_F\Tr(\mathsf L_m A_{{\rm em},m}),\\
 Q_T(\omega)&=Q(\omega)-Q_L(\omega).
 \end{aligned}
 \label{eq:Q}
\end{equation}
These are the main-text spectra: Fig.~2(a) there plots $Q$, and
Fig.~2(b) plots the frequency-dependent ratio $Q_L/Q$.
For those curves, $\alpha=0.2$, $\chi=0,0.6,0.9$, and the retained
branches are continued from Schwarzschild $j\le7/2$.
In particular, $\Tr A_{{\rm em},m}=\sum_b\Gamma_{mb}$, so the total
spectrum follows from the summed greybody factors.
The longitudinal spectrum additionally requires $\mathsf L_m$ and the
outgoing matrix $A_{{\rm em},m}$.
If the reflection matrix is diagonal, this reduces to
\begin{equation}
 \Tr(\mathsf L_m A_{{\rm em},m})
       =\sum_b(\mathsf L_m)_{bb}\Gamma_{mb}.
 \label{eq:diagonalhelicitysum}
\end{equation}
Otherwise the full matrix product in \eqref{eq:Q} retains its
off-diagonal terms.

The mode sum includes all four massive particle spin states. Rates
are given for one Majorana species or one charge sector of a Dirac
species; including Dirac particles and antiparticles doubles them.
The longitudinal fractions are unchanged by that overall factor.
This stationary calculation uses the stated physical-mode Unruh
prescription and does not establish a global quantum-state construction
for the constrained Kerr field.

\begin{figure}[tbp]
 \centering
 \makebox[\columnwidth][c]{\hspace*{-0.17in}\includegraphics[width=3.375in]{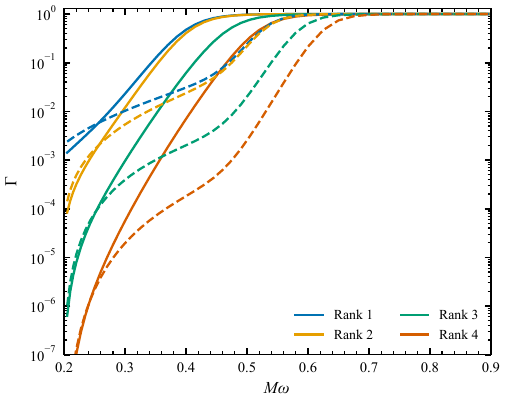}\hspace*{0.17in}}
 \caption{Leading four angular ranks at $m=3/2$, $\alpha=0.2$.
 Solid and dashed curves use $\chi=0.6,0.9$.
 Rank labels a spectral ordering and is not a helicity.}
 \label{fig:greybody}
\end{figure}

For a dimensionless frequency band $\mathcal B_\omega$, define its
longitudinal number fraction $f_{L,\mathcal B_\omega}$ by
\begin{equation}
 f_{L,\mathcal B_\omega}
 =\frac{\int_{M\omega\in\mathcal B_\omega}Q_L\,\dd\omega}
 {\int_{M\omega\in\mathcal B_\omega}Q\,\dd\omega}
 =(0.776,\ 0.444,\ 0.220)\quad
 \text{for }\chi=(0,\ 0.6,\ 0.9).
\end{equation}
The displayed values use the common main-text band
$\mathcal B_\omega=[0.2005,0.6]$ at $\alpha=0.2$.
This integrated fraction weights the plotted ratio $Q_L/Q$ by
$Q\,\dd\omega$; it is not an unweighted frequency average.
The positive and negative helicities are grouped by absolute value.
These fractions are not a net chirality. Fig.~\ref{fig:greybody}
shows the greybody factors of the four leading angular branches at
$m=3/2$ for $\chi=0.6,0.9$.

For a retained band $\mathcal B_\omega$, let
$\dot N_{\mathcal B_\omega}=\int_{M\omega\in\mathcal B_\omega}Q\,\dd\omega$
and $\dot E_{\mathcal B_\omega}
=\int_{M\omega\in\mathcal B_\omega}\omega Q\,\dd\omega$ be its number
and energy rates. A dot denotes a derivative with respect to $t$;
$\dot N=\int_\mu^\infty Q\,\dd\omega$ denotes the full number rate.
Table~\ref{tab:rates} gives these finite-interval estimates on larger bands.
\begin{table}[tbp]
 \caption{Finite-interval particle and energy rates at $\alpha=0.2$,
 with the four-state species convention used in the text.}
 \begin{ruledtabular}
 \begin{tabular}{c c c c c}
 $\chi$&$M\omega$ interval&$M\dot N_{\mathcal B_\omega}$&
 $f_{L,\mathcal B_\omega}$&$M^2\dot E_{\mathcal B_\omega}$\\
 $0$&$[0.2001,1.3]$&$1.0618\,10^{-4}$&$0.7762$&$2.9587\,10^{-5}$\\
 $0.6$&$[0.2001,1.6]$&$5.6471\,10^{-4}$&$0.4402$&$1.9889\,10^{-4}$\\
 $0.9$&$[0.2001,2.8]$&$3.7765\,10^{-3}$&$0.2176$&$1.8777\,10^{-3}$
 \end{tabular}
 \end{ruledtabular}
 \label{tab:rates}
\end{table}
The high-frequency sums reach Schwarzschild-origin $j\le11/2$ at
$\chi=0.6$ and $j\le15/2$ at $\chi=0.9$.
Selected tail checks omit sectors with
$(\omega-m\Omega_H)/T_H>38$. No bound on the infinite mode sum is inferred.

The strip $\mu<\omega<0.2001/M$ is uncomputed. Its width times the
first retained spectral value estimates contributions
$4.8\,10^{-8}$, $1.1\,10^{-7}$, and $2.1\,10^{-7}$ to $M\dot N$;
these are not upper bounds. Points at $M\omega=0.200025$ fail the
flux-orthogonality checks and are excluded.

BlackHawk v2 describes massless spin-$3/2$ Schwarzschild factors and a
mass cutoff prescription \cite{BlackHawk2021}. The v3 Kerr tables also
include spin $3/2$, using the massless Teukolsky equation
\cite{BlackHawk2026}. These are distinct from the full $\mu\ne0$ RS
propagation problem. Finite-mass vector emission provides a related
example of polarization-dependent spectra \cite{Calza2026}.

Let $Q^{(0)}_{3/2}$ be the massless spin-$3/2$ number spectrum with two
helicities, and $\Theta$ the Heaviside step function. The equal-state-count
mass-cutoff reference is
\begin{equation}
 Q_{\rm cut}=2Q^{(0)}_{3/2}\Theta(\omega-\mu).
\end{equation}
Here $Q^{(0)}_{3/2}$ includes two helicities, so $Q_{\rm cut}$ has
four states. This benchmark does not reproduce every implementation
detail of a particular code release.
On the common band, massive-to-cutoff rate ratios are
$1.44$, $0.657$, and $0.515$ for the three rotations.

As a diagnostic one can change the weights $n_F$ while keeping
Schwarzschild transmission traces, or change the transmission traces
while keeping Schwarzschild weights. These hybrid curves isolate
factors and are not spectra of another black-hole geometry.
In the common band, changing only the occupation gives longitudinal
fractions $0.327$ and $0.357$ for the $\chi=0.6,0.9$ weights.
The complete physical results are $0.444$ and $0.220$.
Propagation partly offsets the reweighting in the first case and
enhances the reduction in the second.

\bibliography{ref}